\documentclass{aa}

\usepackage{graphicx}
\usepackage{color}

\usepackage{natbib}
\usepackage{txfonts}

\begin{document}

\title{Cosmological formation and chemical evolution of an elliptical galaxy} 

\titlerunning{Chemical evolution of elliptical galaxies}

\author {E. Colavitti\inst{1,}\thanks{email to: colavitti@oats.inaf.it}
\and A. Pipino \inst{2}
\and F. Matteucci\inst{1,3}}

\authorrunning{E. Colavitti et al.}

\institute{Dipartimento di Astronomia, Universit\'a di Trieste, Via G. B. Tiepolo 11, I-34143 Trieste (TS), Italy 
\and Department of Physics and Astronomy, University of Southern California, Los Angeles, CA 90089-0484, U.S.A.
\and I.N.A.F. Osservatorio Astronomico di Trieste, Via G. B. Tiepolo 11, I-34143 Trieste (TS), Italy}

\date{Received xxxx / Accepted xxxx}

\abstract{}
{We aim at studying the effect of a cosmologically motivated gas infall law for the formation of a massive elliptical galaxy in order to understand its impact on the formation of the spheroids.}
{We replace the empirical infall law of the model by Pipino \& Matteucci with a cosmologically derived infall law for the formation of an elliptical galaxy. We constrast our predictions with observations. We also compare the obtained results with those of Pipino \& Matteucci.} 
{We computed models with and without galactic winds: we found that models without wind predict a too large current SNIa rate. In particular, the cosmological model produces a current SNIa which is about ten times higher than the observed values.  Moreover models without wind predict a large current SNII rate, too large even if compared with the recent GALEX data. The predicted SNII rate for the model with wind, on the other hand, is too low if compared with the star formation histories given by GALEX. Last but not least, the mean value for the [Mg/Fe] ratio in the dominant stellar population of the simulated galaxy, as predicted by the cosmological model, is too low if compared to observations. This is, a very important result indicating that the cosmological infall law is in contrast with the chemical evolution.} 
{A cosmologically derived infall law for an elliptical galaxy cannot reproduce all the chemical constraints given by the observations. The problem resides in the fact that the cosmologically derived infall law implies a slow gas accretion with consequent star formation rate active for a long period. In this situation low [Mg/Fe] ratios are produced for the dominant stellar population in a typical elliptical, at variance with observations.}
{}


\maketitle

\section{Introduction}

Elliptical galaxies show an important uniformity in their photometric and chemical properties. The first proposed scenario of elliptical formation was the monolithic collapse scenario (Larson 1974; Matteucci \& Tornamb\`e 1987; Arimoto \& Yoshii 1987; Chiosi \& Carraro 2002). Ellipticals are assumed to have formed at high redshift as a result of a rapid collapse of a gas cloud. This gas is rapidly converted into stars by means of a very strong burst, followed by a galactic wind powered by supernovae and stellar winds. The wind, carring out the residual gas from the galaxy, inhibites further star formation. Minor episodes of star formation, related to gas accretion from the surrounding medium or interactions with neighbours, are not excluded, although they do not influence the galactic evolution. 

The second proposed scenario for elliptical formation was the hierarchical clustering scenario. Hierarchical semi-analytic models predict that ellipticals are formed by several merging episodes which trigger starbursts and regulate the chemical enrichment of the system (White \& Rees 1978). In this case massive ellipticals form at relatively low redshifts through major mergers between spiral galaxies (e.g. Kauffmann \& White 1993; Kauffmann \& Charlot 1998).

The high-redshift formation of elliptical galaxies is supported by observations showing also an increase in the strength of the metal absorption lines (mass-metallicity relation, e.g. Carollo, Danziger \& Buson 1993; Gonzalez 1993; Davies, Sadler \& Peletier 1993; Trager et al. 1998, 2000b) and a reddening of the stellar light (colour-magnitude relation, e.g. Bower et al. 1992) with the velocity dispersion of the galaxies. In other words, the most massive ellipticals are also the most metal-rich. This was interpreted by Larson (1974) as resulting from the galactic winds that should occur later in the most massive objects, allowing the star formation process to continue for a longer period. In the last years, however, this interpretation has been challenged.  This is because there was increasing evidence that the [Mg/Fe] ratio in the dominant stellar population is larger than zero in the core of bright galaxies (e.g. Faber, Worthey \& Gonzalez 1992), suggesting that the star formation lasted for a period shorter than the time at which the pollution from Type Ia SNe becomes important (Matteucci  1994; Weiss, Peletier \& Matteucci 1995; Tantalo \& Chiosi 2002). Moreover, the [Mg/Fe] ratio in the cores of elliptical galaxies increases with galactic mass (Worthey et al. 1992; Weiss et al. 1995; Kuntschner 2000) indicating that higher mass galaxies are, on average, older, more metal-rich and more $\alpha$-enhanced than lower mass galaxies. Moreover the largest galaxies formed their stars $\sim$ 13 Gyr ago, whereas the mean age of low-mass objects is only $\sim$ 4 Gyr (e.g. Nelan et al., 2005) in a manner resembling the 'inverse wind model' of Matteucci (1994). This is clearly at variance with the scenario of Larson. Other possible solutions to the increase on [Mg/Fe] as a function of galactic mass are a variable initial mass function (IMF) and/or a decrease of the DM content as a function of galactic mass (Matteucci, Ponzone \& Gibson 1998). In Pipino \& Matteucci (2004) the authors showed that the majority of the optical properties of elliptical galaxies can be simultaneously reproduced under the assumption that the formation process, occurring at high redshift (as in the monolithic collapse scenario), is stronger and faster in more massive objects with respect to the less massive ones. On the other hand, evidences favouring the hierarchical scenario are represented by the observed interacting galaxies, ellipticals with disturbed morphologies (e.g. Kormendy \& Djorgovski 1989). However, semi-analytical models based on this framework still fail in reproducing the [Mg/Fe]-mass relation in ellipticals (e.g. Nagashima et al., 2005).

Evidences of very recent (but very modest) star formation in early-type galaxies come from the Galaxy Evolution Explorer (GALEX) data (e.g. Yi et al. 2004, Salim et al. 2005). The sensitivity of GALEX to very low star formation rates ($\sim 10^{-3} M_{\odot} \rm yr^{-1}$, Salim et al. 2005) provide us with more reliable data on recent minor episodes of star formation in ellipticals. Therefore one cannot completely rule out the possibility of late mergers, although they do not represent the main process of galaxy formation as they seem to be limited only to (a fraction of) the low mass objects. Monolithic models, obviously, cannot reproduce recent star formation if there is not late-time accretion of satellites. The cosmological model, instead, has a residual star formation by definition. For this reason it is interesting to compare it with the data by GALEX.

Kaviraj et al. (2008a) showed that the medium value of recent star formation in early-type galaxies, defined by the authors as the mass fraction of stars that form in a galaxy within the last 1 Gyr of look-back time in its rest frame, is $\sim$ 5.5 per cent.    

A residual star formation at low redhshift seems quite common in Brightest Cluster Galaxies (BGCs) hosted in cool-core clusters (Hicks \& Mushotzky 2005, O'Dea et al. 2008, Bildfell et al. 2008). Using Bildfell et al. (2008) sample, Pipino et al. (2008b) demonstrate, \emph{for the first time}, a one-to-one correspondence between blue cores in elliptical galaxies and a UV-enhancement observed using GALEX. However, it should be said that only a small fraction of elliptical galaxies is  showing current star formation. In any case it is interesting to understand if this supposed recent star formation really exists or not and how it can influence the chemical evolution of ellipticals.

Generally, models for the formation of normal ellipticals predict that the star formation rate stopped several Gyrs ago. In fact, besides the models quoted before, Merlin \& Chiosi (2006) found that the star formation in early-type galaxies was completed within 4 Gyr, showing that a strong wind occurred. The same was found by Kawata \& Gibson (2003a). They showed that the star formation stops abruptly at an early epoch due to the galactic wind. In another paper (Kawata \& Gibson 2003b) the authors showed that the radiative cooling becomes more efficient, and thus the gas infall rate increases, with decreasing mass of elliptical galaxies.

With this paper we aim at studying the effect of a cosmologically motivated infall law for the formation of a massive elliptical galaxy in order to understand its impact on the formation of the spheroids. The novelty of our approch to model the formation of an elliptical galaxy consists in the fact that we run a dark matter only cosmological simulation, identifing a posteriori a possible dark matter halo which can host an elliptical galaxy. To do this we use in particular the spin parameter, connected to the angular momentum of a galaxy. Other authors studied the formation and evolution of ellipticals, but with a different approach. Among the others, Thomas, Greggio \& Bender (1999) adopted two different scenarios for the formation of an elliptical galaxy: a fast clumpy collapse and the merger of two spirals similar to the Milky Way. Pipino \& Matteucci (2004), instead, adopted an analitical infall law to study the formation and the chemical evolution of an elliptical galaxy.

The paper is organized as follows: in section 2 we describe the adopted model, analyzing the cosmological simulation and the chemical evolution. Section 3 presents the obtained results and section 4 the conclusions.

\section{The model}
The originality of our work is based on the fact that we use a cosmological simulation in order to derive a mass accretion history for an elliptical galaxy to insert in the chemical evolution model by Pipino \& Matteucci (2004). We use the same cosmological simulation by Colavitti et al. (2008), in which the authors derived four infall laws for spiral galaxies like the Milky Way. In this case we identify several dark matter halos which can host an elliptical galaxy, then deriving their mass accretion histories. After this, we chose a particular DM halo to analyze, looking at his peculiar characteristics (see paragraph 2.1). 

\subsection{The cosmological simulation}
The main aim of our work is to follow the chemical evolution of elliptical galaxies in a cosmological context. To this aim, we run a dark matter-only cosmological simulation, using the public tree-code GADGET2 (Springel 2005), in order to produce and study dark matter halos in which elliptical galaxies can form.

We refer the reader to Colavitti et al. (2008) for the details concerning the cosmological simulation and we describe here only the assumptions made to select the right DM halo.
To identify the DM halos that can host an elliptical galaxy we used selection criteria based on two different characteristics. The first one is the mass of the halo. We want to study a mean elliptical galaxy, so it is important to chose an interval of masses in which we can find the it. The second characteristic is the spin parameter. This parameter is a measure of the angular momentum of the halo and, therefore, we look for a small spin parameter in order to be sure to consider a DM halo which can host an elliptical galaxy. These are the selectioned characteristics:

\begin{itemize}
 
\item mass between $1 \cdot 10^{12} M_{\odot}$ and $5 \cdot 10^{12} M_{\odot}$;

\item spin parameter $\lambda < 0.04$; 

\end{itemize}

We found 22 DM halos compatible with our selection criteria. We assumed that the baryonic matter follows the same accretion pattern as the dark matter and that it represents 19\% (the cosmological baryon fraction) of all the infalling matter. In this way, we obtained the baryon infall law from the mass accretion history of each halo.
Having 22 infall laws and knowing that elliptical galaxies form the bulk of their stars rapidly, we looked for a halo with a large  mass accretion rate at high redshift. This characteristic, in conjunction with a low spin parameter, was found only in one halo, being representative of a typical elliptical galaxy. 

One may argue that this halo is not so representative, being these three characteristics not essential to characterize an elliptical galaxy. However the mass, the spin parameter and the redshift of formation of a galaxy are the same parameter used in Colavitti et al. (2008) to identify a spiral galaxy similar to the Milky Way, where the authors obtained good results.
It is important to underline some shortcomings of our approach and in particular the fact that ellipticals are tipically found in clusters of galaxies, whereas in our simulation we model an isolated elliptical galaxy. In a galaxy cluster there are some important physical processes (i.e. ram-pressure stripping, tidal stripping, harassment among others) that we do not consider in our DM-only simulation and that can influence the properties and the amount of the gas. Another important process that we do not consider is that ellipticals are likely to accrete not only gas during the mergers but also already formed stars. In this case, details on baryons properties become important and they cannot be ignored.

\subsection{Chemical evolution}

Following the approach of Colavitti et al. (2008) we now implement the cosmologically motivated infall law into the Pipino \& Matteucci (2004) (hereafter PM04) chemical evolution model for elliptical galaxies (their Model I).
This model is an updated version of the multizone model of Martinelli et al. (1998) and Pipino et al. (2002). In the model the ellipticals form by fast accretion of gas at high redshift, and the infall law is calibrated to obtain, in absence of galactic winds, a luminous mass of roughly $10^{11} M_{\odot}$ ($1.32 \cdot 10^{11}$ in Table 1). If instead the occurrence of a galactic wind is allowed, then the final luminous mass is slightly smaller ($1.08 \cdot 10^{11}$).

For the star formation rate $\psi$ they adopted the following law:

\begin{equation}
\psi(t) = \nu \cdot \frac{\rho_{gas}(t)}{\rho_{gas}(0)}.
\end{equation}

The constant $\nu$, representing the star formation efficiency, is an increasing function of the galactic mass in order to reproduce the `inverse wind model' (Matteucci 1994; Matteucci et al. 1998).

For Type Ia SNe they assumed a progenitor model made of a C-O white dwarf plus a red giant (Whelan \& Iben 1973). The resulting rate is:

\begin{equation}
R_{SNIa} = A \int_{M_{Bm}}^{M_{BM}} \psi(M_{B}) \int_{\mu_{m}}^{0.5} f(\mu)\psi(t-\tau_{M2}) \; d\mu \; dM_{B}.
\end{equation} 

(Greggio \& Renzini 1983; Matteucci \& Greggio 1986), where $M_{B}$ is the total mass of the binary system, $M_{Bm} = 3 M_{\odot}$ and $M_{BM} = 16 M_{\odot}$ are the minimum and maximum masses allowed for the adopted progenitor systems, respectively. $\mu = M_{2}/M_{B}$ is the mass fraction of the secondary and $\mu_{m}$ is its minimum value. The constant $A$ represents the fraction of binary systems in the IMF which are able to give rise to SNIa explosions. 

For the infall law PM04 adopted the expression:

\begin{equation}
\Big[\frac{dG_{i}(t)}{dt}\Big]_{infall} = X_{i,infall} \; C \; e^{-t/\tau}.
\end{equation} 

where $X_{i,infall}$ describes the chemical composition of the accreted gas, assumed to be primordial, and $\tau$ is the infall time-scale. $C$ is a constant obtained by integrating the infall law over time and requiring that $\sim$ 90\% of the initial gas has been accreted at the time when the galactic wind occurs. 

Differently from PM04, here we use the yields by Woosley \& Weaver (1995).

We study three different models: Model 1 is the same model as in Pipino \& Matteucci (2004), with the difference that we on purpose turned off the galactic wind (a natural consequence of SN feedback), in order to have a residual star formation rate at low redshift; Model 2, instead, is exactly the same model as in Pipino \& Matteucci (2004). The wind occurs after 0.92 Gyr from the beginning of the formation of the galaxy, when the thermal energy of the gas heated by SNe (II and Ia) equates the binding energy of the gas, thus halting the star formation. Model 3 is our cosmological model. In this case, we maintained all the characteristics of Model 1, but changing the infall law. In this case we do not have galactic wind since we want to have a residual present time star formation rate.  

In Table 1 we show, for all the models,  the accreted luminous mass (see comments above) and  the star formation efficiency (i.e. the star formation rate per unit mass of gas) and the infall timescale, the same appearing in eq. (3). In the last column we show the time of galaxy formation, this  timescale can be defined as the time at which the galaxy accreted half of its mass. As one can see, for the monolithic models $\tau \sim t_{form}$. Therefore, one can note that for the Models 1 and 2 this time corresponds to a redshift $z\sim$5.0, whereas for Model 3 is $z\sim$ 2.0. The meanings of the parameters are the same as in PM04. 

\begin{table*}
\caption{Luminous mass, star formation efficiency and time of formation for all the models.}
\centering
\begin{tabular}{c|c|c|c|c}
\noalign{\smallskip}
\hline
\hline
\noalign{\smallskip}
Model & $M_{lum}$ [$M_{\odot}$] & $\nu$ & $\tau$ [Gyr] & $t_{form}$ [Gyr] \\
\noalign{\smallskip}
\hline
\noalign{\smallskip}
1 & $1.32 \cdot 10^{11}$ & 10 & 0.4 & 0.4 \\
2 & $1.08 \cdot 10^{11}$ & 10 & 0.4 & 0.3\\
3 & $1.32 \cdot 10^{11}$ & 10 &     & 4.0 \\
\noalign{\smallskip}
\hline
\hline
\end{tabular}
\end{table*}

\section{Results}

In this section we present our results. In particular, in the first subsection we show several plots where we compare the three models. In the upper panels of each plot it can be seen Model 1, i.e. the model by Pipino \& Matteucci (2004) without wind. In the middle panels we show the same model, but with the wind (Model 2). In the bottom panel it can be seen Model 3, i.e. the cosmological model without wind. We show the different infall laws and total masses, the star formation rate and the SNIa rate, the behavior of the [Fe/H] ratio as a function of time and, finally, the [O/Fe] and [Mg/Fe] as a function of [Fe/H]. 

In the second subsection we present some comparisons of our results with observable data given by Thomas et al. (2005), Mannucci et al. (2008) and Kaviraj (2008b). In particular, we show the mean values for [Fe/H], [Mg/H] and [Mg/Fe] for each model, the current SNIa and SNII rates and the mass in stars formed in the last 1.0 and 6.2 Gyr by each model.   

\subsection{Comparison between the three models}

\begin{figure}
\centering
\includegraphics[width=0.45\textwidth]{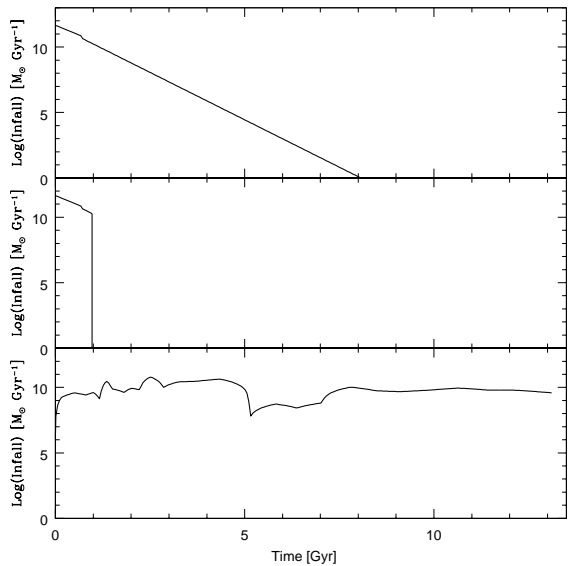}
\caption[]{The gas infall  rate vs time for the three models. Upper panel: Model 1; Middle panel: Model 2; Bottom panel: Model 3.}
\label{inf}
\end{figure}

Figure \ref{inf} shows the infall laws for all the models as a function of time. The infall law of Model 1 and 2 has a peak at high redshift, and lasts for 1 Gyr. In particular, the infall law of Model 2 becomes equal to zero after the peak, since the wind starts. The infall law of Model 3, as we know, is cosmologically derived. It does not present a peak, having an important accretion episode between 2 and 5 Gyr and several minor episodes during the whole formation of the galaxy.  

\begin{figure}
\centering
\includegraphics[width=0.45\textwidth]{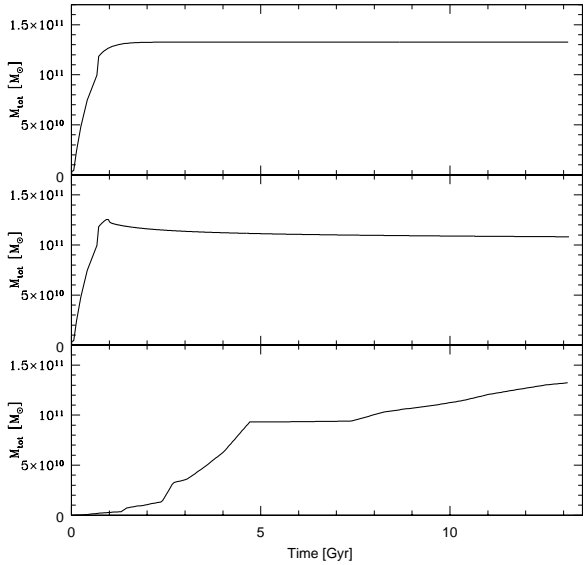}
\caption[]{Total mass vs time for the three models. Upper panel: Model 1; Middle panel: Model 2; Bottom panel: Model 3.}
\label{mas}
\end{figure}

In Figure \ref{mas} it can be seen the total mass of all the models as a function of time. The mass of Model 1 increases during the first Gyr of formation, remaining constant when the infall law rapidly decreases. Model 2 has the same behavior during the first Gyr, whereas it slightly decreases after 1 Gyr, since the wind starts and part of the gas is ejected. The mass accretion history of Model 3, instead, is very different from those of Model 1 and 2. It increases very slowly until $\sim$ 2.5 Gyr, when the slope increases. Then, between $\sim$ 4.8 and $\sim$ 7.5 Gyr it remains more or less constant, starting to increase again up to the end of the formation of the galaxy.      

\begin{figure}
\centering
\includegraphics[width=0.45\textwidth]{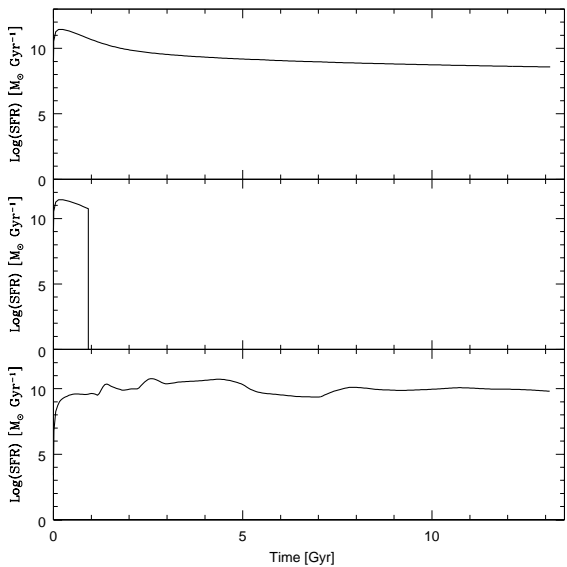}
\caption[]{The star formation rate (SFR) vs time for the three models. Upper panel: Model 1; Middle panel: Model 2; Bottom panel: Model 3.}
\label{sfr}
\end{figure}

Figure \ref{sfr} shows the star formation rate as a function of time for the three models. It can be seen that it is very similar to the infall rate of each model. The difference between Model 1 and 2 is that, when the wind starts, the SFR of Model 2 suddenly becomes equal to zero, whereas in Model 1 it goes to zero very slowly. In Model 3 the SFR never stops, since in this case there is no wind and the infall rate is always different from zero.  

\begin{figure}
\centering
\includegraphics[width=0.45\textwidth]{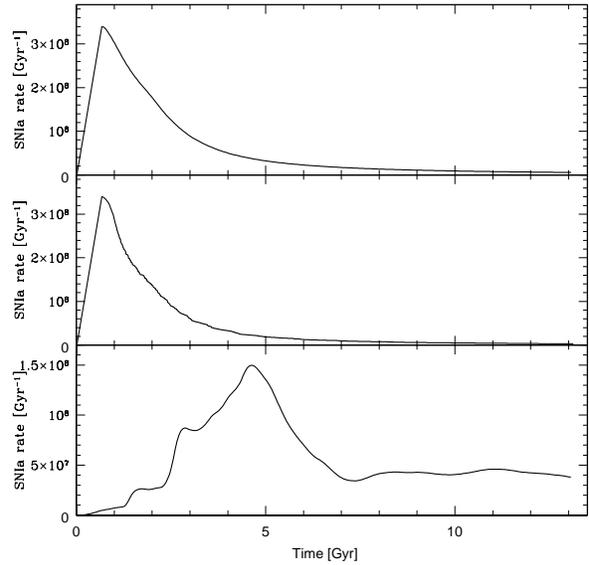}
\caption[]{The SNIa rate vs time for the three models. Upper panel: Model 1; Middle panel: Model 2; Bottom panel: Model 3.}
\label{snia}
\end{figure}

In Figure \ref{snia} it can be seen the SNIa rate as a function of time. The behavior of Model 1 and 2 are very similar, even if the SNIa rate of Model 2 decreases more rapidly. Model 3 produces a large amount of SNIa between 2 and 7 Gyr, when the infall rate is larger. After this interval the SNIa rate becomes nearly constant until the end of the evolution. We do not present a plot for SNII because, being the life of their progenitor very brief, their behavior traces out the SFR. 

\begin{figure}
\centering
\includegraphics[width=0.45\textwidth]{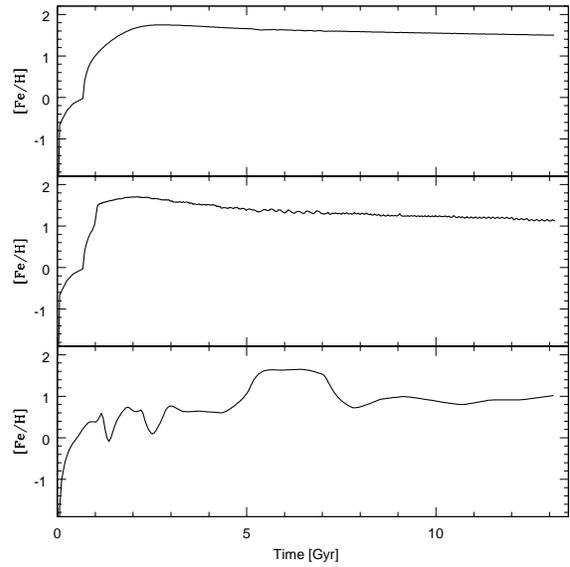}
\caption[]{[Fe/H] vs time for the three models. Upper panel: Model 1; Middle panel: Model 2; Bottom panel: Model 3.}
\label{feh}
\end{figure}

Figure \ref{feh} presents the [Fe/H] as a function of time for all the models. Both Model 1 and Model 2 reach the largest value for the [Fe/H] ratio after about 2 Gyr from the beginning of the formation. After this peak the [Fe/H] slightly decreases, in particular in Model 2 where, at the end, [Fe/H] $\sim$ 1.2. In Model 3 the [Fe/H] ratio has a behavior very different form that of Model 1 and 2. In the first three Gyrs the [Fe/H] has several peaks and depressions. The latter are due to the two peaks in the infall rate, at 1.5 and 2.5 Gyr which dilute the Fe abundance because the infalling gas is assumed to have a primordial chemical composition (no metals). 
After the first peaks the [Fe/H] ratio increases rapidly, then becoming constant for more or less two Gyr, decreasing again and then keeping constant and  equal to $\sim$ 1.0 dex at the end of the 
present time.  

\begin{figure}
\centering
\includegraphics[width=0.45\textwidth]{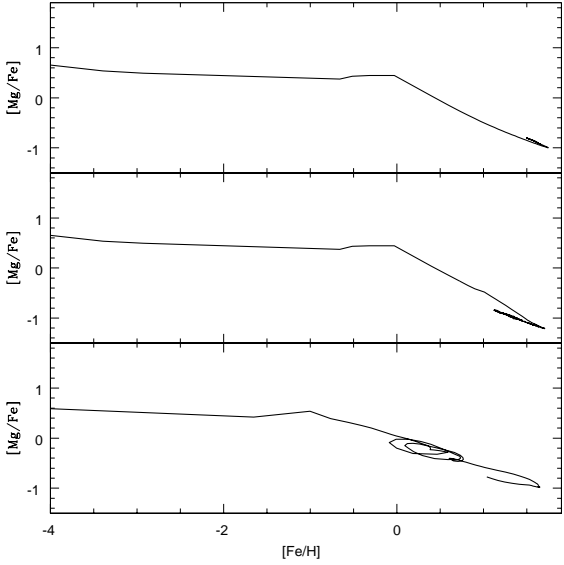}
\caption[]{[Mg/Fe] vs [Fe/H] for the three models. Upper panel: Model 1; Middle panel: Model 2; Bottom panel: Model 3.}
\label{mgfe}
\end{figure}

In Figure \ref{mgfe} it can be seen the [Mg/Fe] as a function of [Fe/H]. All the three models initially have the same behavior. However the [Mg/Fe] of Model 3 starts to decrease at lower metallicities than in Model 1 and 2. In Model 3 the changing in slope happens at [Fe/H] $\sim$ -1, whereas in Model 1 and 2 it happens at [Fe/H] $\sim$ 0. In Model 1 and 2 the [Mg/Fe] becomes equal to zero at [Fe/H] $\sim$ 0.4, whereas in Model 3 the [Mg/Fe] ratio becomes zero when the [Fe/H] $\sim$ 0.1. This is due to the longer and less efficient star formation predicted by Model 3. In fact, different star formation histories produce different [$\alpha$/Fe] abundance patterns, favoring high values for a large interval of [Fe/H] in regimes of high and short star formation (Matteuccci, 2001). The high and short star formation rate produced [Mg/Fe] ratios in excellent agreement with those measured in the Bulge (Ballero et al. 2007) and in ellipticals(Pipino \& Matteucci 2004).

\subsection{Comparison with observables}
 
In order to study the mean properties of the stellar component in ellipticals, we need to average quantities related to the mean abundance pattern of the stars, which, in turn, can allow a comparison with the observed integrated spectra\footnote{We recall that in elliptical galaxies it is not possible to resolve single stars and, hence, determine the stellar metallicity distribution.}. To this scope, we recall that both real and model galaxies are made of CSP, namely a mixture of several SSP, differing in age and chemical composition according to the galactic chemical enrichment history, weighted with the SF rate. On the other hand, the observed line-strength indices are usually tabulated only for SSPs as functions of their age, metallicity and (possibly) $\alpha$-enhancement. In particular we make use of the mass-weighted mean stellar metallicity as defined by Pagel \& Patchett (1975, see also Matteucci 1994): 

\begin{equation}
<Z> = {1\over S_f} \int_0^{S_f} Z(S) \; dS.
\label{PP75original}
\end{equation}

where $S_f$ is the total mass of stars ever born contributing to the light at the present time and Z is the metal abundance (by mass) in the gas out of which an amout of stars $S$ formed. In practice, we make use of the stellar mass distribution as a function of Z in order to derive the mean metallicity in stars.

In Table 2 we show the mean value for [Fe/H], [Mg/H] and [Mg/Fe] for all the models. The mean values for Model 1 and Model 3 are taken during the whole lifetime of the galaxy, while in Model 2 the mean values are representative for the whole period prior to the wind. It can be seen that the [Mg/Fe] value for Model 3 is too low and it is in contrast with the values given by Thomas et al. (2005), where for an elliptical galaxy of the same mass the value of the fit of the observed data is $\sim$ 0.2 dex

\begin{table*}
\caption{Mean values for [Fe/H], [Mg/H] and [Mg/Fe] for all the models. The time for Model 2 is the time at which the wind occurs.}
\centering
\begin{tabular}{c|c|c|c|c}
\noalign{\smallskip}
\hline
\hline
\noalign{\smallskip}
Model & Time [Gyr] & [Fe/H] & [Mg/H] & [Mg/Fe] \\
\noalign{\smallskip}
\hline
\noalign{\smallskip}
1 & 13.12 & 0.677 & 0.358 & 0.434 \\
2 & 0.92 & -0.078 & 0.270 & 0.511 \\
3 & 13.12 & 0.777 & 0.299 & -0.304 \\
\noalign{\smallskip}
\hline
\hline
\end{tabular}
\end{table*} 

However, Thomas et al. (2004) found that a non-negligible fraction of low-mass ellipticals in their sample shows evidences of a young (i.e. age $< 2$ Gyr) and metal-rich ($[Z/H]\sim0.2$) stellar population which represents $\sim 10-20 \%$ of their total stellar mass. These values are independent from the environment and the typical [$\alpha$/Fe] of this population are $\sim$ 0.2 dex lower than the bulk of the stars, as expected if the young component formed out of gas strongly polluted by SNIa. 

As seen in the introduction, evidences of very recent (but modest) star formation in early-type galaxies come from the GALEX data. In Table 3 we show the mass in stars formed in the last 1 and 6.2 Gyr by the three models, comparing it with the total amount of stars and with the observed values by Kaviraj (2008b). We can see that Model 2 does not form stars in the last 6.2 Gyr. These is due to the presence of a wind which stops the star formation at early times. Model 3 predicts very well the mass of stars formed in the last Gyr, whereas it produce less stars than expected in the last 6.2 Gyr. The residual star formation of Model 1, instead, cannot reproduce neither the amount of stars formed in the last Gyr (according to Galex) nor those formed in the last 6.2 Gyr, being the obtained values too low. We recall, however, that elliptical galaxies with some residual recent star formation are only a minor fraction of the whole class of ellipticals; therefore the disagreement between Model 1 and the GALEX observations is not emphasizing any model deficiency.

\begin{table*}
\caption{Mass in stars formed in the last 1.0 and 6.2 Gyr for all the models compared with the total amount of stars. Observed values by Kaviraj (2008b).}
\centering
\begin{tabular}{c|c|c|c}
\noalign{\smallskip}
\hline
\hline
\noalign{\smallskip}
Model & $M_{*}$ [$M_{\odot}$] (last 1.0 Gyr [z = 0.08]) & $M_{*}$ [$M_{\odot}$] (last 6.2 Gyr [z = 0.7]) & $M_{*}$ [$M_{\odot}$] (13.12 Gyr) \\
\noalign{\smallskip}
\hline
\noalign{\smallskip}
1 & 4.17 $\cdot 10^{8}$ (0.21\%) & 3.66 $\cdot 10^{9}$ (1.85\%) & 1.98 $\cdot 10^{11}$ \\
2 & 0.00 (0.00\%) & 0.00 (0.00\%) & 1.58 $\cdot 10^{11}$ \\
3 & 7.95 $\cdot 10^{9}$ (4.25\%) & 5.54 $\cdot 10^{10}$ (29.63\%) & 1.87 $\cdot 10^{11}$ \\
\noalign{\smallskip}
\hline
\hline
Kaviraj (2008b) & 1 - 6 \% & 5 - 13 \% & \\
\noalign{\smallskip}
\hline
\hline
\end{tabular}
\end{table*} 

Table 4 shows the current SNIa and SNII rate for the three models. They are compared with the observed values by Mannucci et al. (2008) and they are in units of SNuM (SN per century per $10^{10} M_{\odot}$ of stellar mass). We can see that Model 3 predicts too large an amount of SNIa and SNII. In particular the value of SNIa for Model 3 is at $\sim$ 32 standard deviations from the observed mean. 

As seen in Table 4, the SN rates for Model 3 are too high with respect to the observed data given by Mannucci et al. (2008). In Table 3 we can see that this is due to the residual star formation of Model 3, not seen in the other two models (especially in Model 2), but confirmed by GALEX. Therefore we find a remarkable lack of consistency between the star formation rates inferred by GALEX and the limits on the the SN rates. In fact, in principle we should be able to observe SNII events if the star formation is still on-going. However, the lower brightness of SNII, combined with the fact that the recent star formation takes place in the the bright centres of the galaxies, implies that the probability of finding SNII with current surveys is probably quite low (Mannucci et al. 2008).

\begin{table*}
\caption{Current SNIa rate and SNII rate in SNum for the three models. Observed values by Mannucci et al. (2008).}
\centering
\begin{tabular}{c|c|c}
\noalign{\smallskip}
\hline
\hline
\noalign{\smallskip}
Model & SNIa rate [SNuM] & SNII rate [SNuM] \\
\noalign{\smallskip}
\hline
\noalign{\smallskip}
1 & 0.060 & 0.028 \\
2 & 0.046 & 0.000 \\
3 & 0.673 & 0.767 \\
\noalign{\smallskip}
\hline
\hline
Mannucci et al. (2008) & $0.058^{+0.019}_{-0.015}$ & $<0.017$ \\
\hline
\hline
\end{tabular}

\end{table*}

\section{Conclusions}

In this paper we studied the effects of a cosmologically derived gas infall law, for the formation of an elliptical galaxy, on several observational constraints pertaining to the chemical evolution. We compare the results of our cosmological model with the results of the model by Pipino \& Matteucci (2004) with and without galactic wind. The conclusions are the following:

\begin{itemize}

\item Models without galactic wind, in which the star formation never stops, predict a too large current SNIa rate. In particular, the cosmological model produces a current SNIa which is about ten times higher than the observed values given by Mannucci et al. (2008) for ellipticals. In order to reproduce the observed SNIa rate with the cosmological model, one should adopt, in the calculation of the SN Ia rate, a value of the constant $A$ (see eq. 2)ten times lower, an unrealistically low value which would predict a too low Type Ia SN rate in the solar vicinity.

\item Models without wind predict also a large current SNII rate. In particular,the cosmological model give a current SNII rate which is higher than the SNIa rate, whereas from observations given by Mannucci et al. (2008) it is known that the SNIa rate is at least three time higher than the SNII rate in ellipticals. On the other hand, the Models 1 and 2, based on the monolithic scenario,  predict almost no current star formation rate and therefore  a negligible number of current Type II SNe, at variance with GALEX indicating recent star formation, although in a small fraction of ellipticals. We have compared our predicted current Type II SN rate with the data by Mannucci et al. (2008). In this case Model 3 predicts far too many Type II SNe whereas Model 1 predicts an acceptable value. 

\item The mean value for the [Mg/Fe] ratio in the dominant stellar population, as predicted by our cosmological model, which is an important indicator for the robustness of a model, is too low, as expected. We predict  [Mg/Fe]$\sim -0.3$ dex. For an elliptical galaxy of the same mass Thomas et al. (2005) give a value of $\sim$ +0.2 dex, corresponding to a gap of about 0.5 dex. This is the most important result, since it suggests that a history of star formation as the one derived by cosmological simulations cannot produce values of the [Mg/Fe] ratio in ellipticals in agreement with observations.  However, it is important to note that our approach has been very simple and that our simulation refers to an isolated galaxy, whereas most of ellipticals reside in clusters. By ignoring that, we might have overlooked some important physical processes such as mergers of stellar systems and all the processes related to the environment (i.e. ram-pressure stripping, tidal stripping and harassment).

\end{itemize} 

%In conclusion, it is important to underline some shortcomings of our approach and in particular the fact that ellipticals are tipically found in clusters of galaxies, whereas in our simulation we model an isolated elliptical galaxy. In a galaxy cluster there are some important physical processes (i.e. ram-pressure stripping, tidal stripping, harassment among others) that we do not consider in our DM-only simulation and that can influence the properties and the amount of the gas. Another important process that we do not consider is that ellipticals are likely to accrete not only gas during the mergers but also already formed stars. In this case, details on baryons properties become important and they cannot be ignored.

%What we can say here is that using a very simple cosmologically derived infall law for an elliptical galaxy it is not possible to reproduce the most important chemical constraints given by observations. In our opinion, the problem resides in the too long and poorly efficient star formation rate predicted for typical and large ellipticals in the hierarchical galaxy formation scenario.

\section{Acknowledgments}

We thank Giuseppe Murante for helpful suggestions and enlightening discussions. E. C. and F. M. 
acknowledge financial support from the Italian Ministry of Research through PRIN2007, N.2007JJC53X-001.

\label{lastpage}

\end{document}